\documentstyle[12pt,aaspp4,flushrt,mystyle]{article}

\newcommand{\nc}{\newcommand}
\nc{\bec}{\begin{center}}
\nc{\enc}{\end{center}}
\nc{\beq}{\begin{equation}}
\nc{\enq}{\end{equation}}
\nc{\beqar}{\begin{eqnarray}}
\nc{\enqar}{\end{eqnarray}}
\nc{\bei}{\begin{itemize}}
\nc{\eni}{\end{itemize}}
\nc{\bee}{\begin{enumerate}}
\nc{\ene}{\end{enumerate}}

\nc{\lig}{light curve}
\nc{\namely}{{\it viz.}}
\nc{\ea}{et al.\ }

\nc{\pl}{period--luminosity}
\nc{\plr}{period--luminosity relation}
\nc{\plc}{period--luminosity--color}
\nc{\pca}{period--color--amplitude}

\nc{\rv}{R_V}
\nc{\av}{A_V}
\nc{\ai}{A_I}
\nc{\lp}{\log (P)}
\nc{\teff}{T_{\mathrm{eff}}}
\nc{\gteff}{\mbox{$ \log \teff \,$--$\,\log g $}}
\nc{\chisq}{\chi ^2}
\nc{\msun}{${\mathcal M}_\odot$}
\nc{\ksm}{km\,s$^{-1}$\,Mpc$^{-1}$}

\nc{\BV}{(B-V)}
\nc{\UB}{(U-B)}
\nc{\vi}{(V-I)}
\nc{\eBV}{E(B-V)}
\nc{\eUB}{E(U-B)}
\nc{\evi}{E(V-I)}
\nc{\cBV}{(B-V)_0}
\nc{\cUB}{(U-B)_0}
\nc{\cvi}{(V-I)_0}

\nc{\vmax}{V_{\mathrm{max}}}
\nc{\vmin}{V_{\mathrm{min}}}
\nc{\uncv}{\langle V\rangle }
\nc{\unci}{\langle  I\rangle }
\nc{\cv}{\langle  V\rangle _0}
\nc{\ci}{\langle  I\rangle _0}
\nc{\uncBVav}{\langle  B-V\rangle }
\nc{\uncviav}{\langle  V-I\rangle }
\nc{\viav}{\langle  V-I\rangle _0}
\nc{\BVav}{\langle  B-V\rangle _0}
\nc{\UBav}{\langle  U-B\rangle _0}
\nc{\uncvimax}{(V-I)|_{\mathrm{at}\: V_{\mathrm{max}}}}
\nc{\uncvimin}{(V-I)|_{\mathrm{at}\: V_{\mathrm{min}}}}
\nc{\vimax}{(V-I)_0|_{\mathrm{at}\: V_{\mathrm{max}}}}
\nc{\vimin}{(V-I)_0|_{\mathrm{at}\: V_{\mathrm{min}}}}
\nc{\viamp}{\Delta (V-I)}
\nc{\vamp}{\Delta V}
\nc{\visyn}{\vi_{\mathrm{synth}}}

\nc{\phiz}{\phi_0}
\nc{\phimin}{\phi_{\mathrm{min}}}
\nc{\phimax}{\phi_{\mathrm{max}}}
\nc{\phinew}{\phi_{\mathrm{n}}}
\nc{\phitilnew}{\phi'}

\lefthead{Narasimha \& Mazumdar}
\righthead{Cepheid Distance to M100}

\begin{document}

\title{Cepheid Distance to M100 in Virgo Cluster}

\author{D. Narasimha\altaffilmark{1}
        and Anwesh Mazumdar\altaffilmark{2}}
\affil{Department of Astronomy and Astrophysics,
       Tata Institute of Fundamental Research, Homi Bhabha Road,
       Mumbai 400005, India.}

\altaffiltext{1}{e-mail: dna@astro.tifr.res.in}
\altaffiltext{2}{e-mail: anwesh@astro.tifr.res.in}

\begin{abstract}
A measurement of the distance to Virgo Cluster by a direct method
along with a realistic error analysis is important for a reliable
determination of the value of Hubble Constant. Cepheid variables 
in the face-on spiral M100 in the Virgo Cluster were observed
with the Hubble Space Telescope in 1994
under the HST Key Project on the Extragalactic Distance Scale.
This work is a reanalysis of the HST data following our study of 
the Galactic Cepheids (in an accompanying communication).
The periods of the Cepheids are determined using two independent methods and 
the reasons for varying estimates are analyzed. The 
log(period) vs $V$-magnitude relation is re-calibrated using LMC data as well as
available HST observations for  three  galaxies  and the slope
is found to be $-3.45 \pm 0.15$.
A prescription to compute  correction for the flux-limited incompleteness 
of the sample is given and a correction of 0 to 0.28 magnitude in
$V$-magnitude for Cepheids in the period range
of 35 to 45 days is applied. The extinction correction is carried out
using {\em period vs mean $\vi $ color} and {\em $V$-amplitude vs $\vi $
color at the brightest phase} relations. The distance to M100 is 
estimated to be $20.4 \pm 1.7$ (random) $\pm 2.4$ (systematic) Mpc.
\end{abstract}


\section{Introduction}
\label{sec:intro}

A natural scale length for the Universe is provided by the Hubble Constant 
($H_0$) and undoubtedly a  determination of its value is one of the fundamental
problems of cosmology.  Over the years, there has been much debate about 
the value of $H_0$ and the present estimates range from less than 50 \ksm\ 
to over 80 \ksm. Most probably the major reason for the discrepancy is
the conventional distance ladder involving multiple steps. Its main
drawback is that analysis of the systematic errors becomes difficult
when the calibrating local sample and the observed sample at the next
step of the ladder are not identical. Consequently, it is believed that
an accurate measurement of the distance to a galaxy cluster which is 
$\sim $ 20--30 Mpc away, without involving intermediate steps, will lead 
to a reliable direct estimate
of the value of $H_0$, provided the recession velocity of the cluster is
independently known. The Virgo Cluster, which is our nearest cluster of
galaxies, is fairly rich in terms of galaxy population, and an average of
the distances to the individual galaxies by different methods would provide
a good estimate to its mean distance.

One of the key projects of the Hubble Space Telescope (HST) was devoted to the
calibration of the extragalactic distance scale 
for a determination of $H_0$ with reasonable accuracy. 
An examination of the systematic errors in the Cepheid \plr\ and 
measurement of the distance to the Virgo
Cluster through Cepheid observation were among the primary aims of this key 
project. The nearly face-on spiral M100 in the Virgo Cluster was observed
on 12 epochs over a span of $\sim$ 57 days in 1994 with the HST using
the filters F555W and F814W, which are almost equivalent to the Johnson V
and Cousins I bands respectively (\cite{freedman:94}). 
The advantage of choosing this particular galaxy is that being nearly face-on, 
the errors due to extinction and reddening are expected to be minimal, and 
further, it is considered to be very similar
to the Milky Way in terms of age, chemical composition etc. However, its
position relative to the center of the Virgo Cluster is not known 
accurately, and that introduces some uncertainty in the Virgo distance
derived from direct distance estimation to M100.
Ferrarese \ea (1996) reported observations of 70 Cepheids in M100, and 
obtained a distance of $ 16.1 \pm 1.3$ Mpc. The value of the Hubble Constant
was calculated to be $88 \pm 24$ \ksm. On the other hand, Sandage and 
collaborators re-calibrated the distance to a few galaxies,
where supernovae of type Ia were detected earlier, by observing the Cepheids
in those galaxies with the HST. They obtained a mean absolute $B$ magnitude
at peak of $-19.6$ for normal SN Ia and consequently, a value of $ 52 \pm 9$
\ksm\ for the Hubble Constant (\cite{sandage:94}; \cite{saha:94}).
However, more recent publications indicate a better reconciliation in the
value of $H_0$. Freedman \ea (1998) summarize a value of 
$73 \pm 6$ (statistical) $\pm 8$ (systematic) \ksm, as compared to 
$55 \pm 3$ (internal) \ksm\ quoted by Sandage's group (\cite{saha:96}).

The present work is a re-analysis
of the HST data on M100 Cepheids, based on a general calibration of 
Galactic Cepheids, presented in a companion paper (which we refer henceforth
as Paper I). The specific problems we address here are the following:
\bei
\item
Period--Luminosity relation applicable to the Cepheids of period $\ga $ 15
days generally observed in distant galaxies.
\item
Importance of the incompleteness correction and quantification of the effect.
\item
Uncertainty in the periods of the Cepheids in M100 due to the
phase sampling techniques applied as well as the large error in $V$-magnitude, 
particularly at low flux levels.
\eni

The central idea behind distance measurement with Cepheids is the \plr.
However, the values of both the slope and the intercept of this relation
continue to be subjects of lively debate. There appears to be a distinct 
difference in the value of the slope between Cepheids of low and high periods. 
While applying the \plr\ to distant galaxy samples, where only higher period
Cepheids can be detected due to flux limitation, this distinction becomes even
more crucial. We address this question on the basis of our study of 
Galactic Cepheids (Paper I) which demonstrates a clear division between
two classes of Cepheids, one with periods $\leq 15$ days, 
the other at higher periods.
The zero-point of the \plr\ is another quantity which needs to be fixed
unambiguously in order to obtain reliable estimates of distance. We use the
recent calibration of the local Cepheids by the Hipparcos mission 
(\cite{fc:97}), rather than the distance to the Large Magellanic Cloud which is normally treated as the calibrating point for the distance scale.

A crucial aspect of our new analysis is the correction for incompleteness
of the Cepheid sample. Since the Cepheid \plr\ has an intrinsic scatter
due to the finite width of the instability strip at a given period, 
the Cepheids are observed to be spread over a range of luminosities.
All the observed Cepheids in M100 have $V$-magnitudes between 24 and 27
mag. At such faint flux levels it is very likely that for a fixed period,
the fainter Cepheids would escape detection, and only the brighter ones
will appear in the surveys. 
This selection bias would have a systematic effect on the
period--$V$-magnitude slope, especially at low periods, reminiscent of the
Malmquist bias discussed in the literature. In order to counter 
this effect, one has to take into account the undetected Cepheids, which
can be done by adequately correcting the observed magnitudes to fainter 
levels. Obviously, the amount of correction depends on the scatter of the
\pl\ diagram. We have devised a formalism to correct for this incompleteness
effect which we demonstrate to be present to a large extent in the M100
sample.

We have tried to estimate the correction for extinction and
reddening, which again, is based on our study of Galactic Cepheids (Paper I).
However, in the absence of multi-wavelength observations, this treatment
is rather limited, and is based on \pca\ relations of Cepheids. Also, 
for the same reason we could not isolate the extinction correction from the
incompleteness correction which ideally we should have been able to. 

This paper is organized as follows. In Section~\ref{sec:per_det} we present
our methods of determination of Cepheid periods and photometric parameters.
The question of choosing the correct \plr\ is addressed in 
Section~\ref{sec:plr}. In Section~\ref{sec:incomp}, we devise a formalism 
for the incompleteness correction of a biased Cepheid sample, and the 
mathematical aspects of compensation for flux-limited bias are described
in the Appendix. 
Section~\ref{sec:extcor} deals with the reddening and extinction corrections 
and the essentials of the numerical methods. The results and major 
contributions to errors are discussed in Section~\ref{sec:results} and
some remarks on the conclusions are presented in Section~\ref{sec:summary}.

\section{Determination of Period, Color and Magnitude}
\label{sec:per_det}

Ferrarese \ea (1996) have tabulated the observed periods as well as
magnitudes after conversion to an equivalent V and I band for the 70
Cepheids in M100 observed by the HST in 1994. For each Cepheid,
there are at most 12 V and 4 I band data, very often only 11 in V and 3 in I
band that are useful for the analysis. We estimate from an inspection of the 
HST data, that the signal to noise
ratio for the F555W filter (V-band) data is typically 6 to 8 for low
period Cepheid variables and 10 to 15 for the higher period ones. The
signal to noise ratio for the F814W filter is in general worse, but the
data turns out to be useful when we have to discriminate between two 
independent estimates of the period computed from the V-band data.

A determination of the period and amplitude from sparsely sampled data with 
inadequate signal to noise ratio does not generally yield a unique result.
According to Bhat, Gandhi and Narasimha (1998), a reasonable criterion to
obtain the period approximately $98\%$ of the times to
better than $2\%$ accuracy may be stated as follows: 
If the signal to noise ratio (SNR) is better than 30 and there are two 
independent sine components in the signal, 
13 optimally spaced sampling points are sufficient to extract the signal,
but if there are three components we require 15 data points. For 
SNR decreasing from around 30 to 10, the requirement increases approximately
linearly with the inverse of SNR from 15 to 22 samplings
for a three-component signal
and thereafter the requirement appear to increase more rapidly to 36
data points when SNR is 3. Naturally, the HST data on M100 could yield
somewhat differing periods and amplitude of pulsation depending on the
method employed. We decided to recompute the periods by two independent
methods for comparison with the periods given by Ferrarese \ea (1996).

We used a modified version of the period-searching program by
Horne and Baliunas (1986) based on the method due to Press and Rybicki (1989).
When our derived periods were substantially different from the value
obtained by the HST Group, we used a template for the V and I band \lig s
prepared on the basis of the Galactic Cepheid variables (cf.\ Paper I)
to get another estimation for the pulsation period. Our final periods
agree with those of Ferrarese et al.\ to within $10\%$ in general,
but at high periods, in some cases the discrepancy is higher. 
We present a comparison of the \lig s of a typical M100 Cepheid, C38
obtained by us with the one derived by the HST group
in Figure~\ref{fig:ligc_comp}.

\figtwo{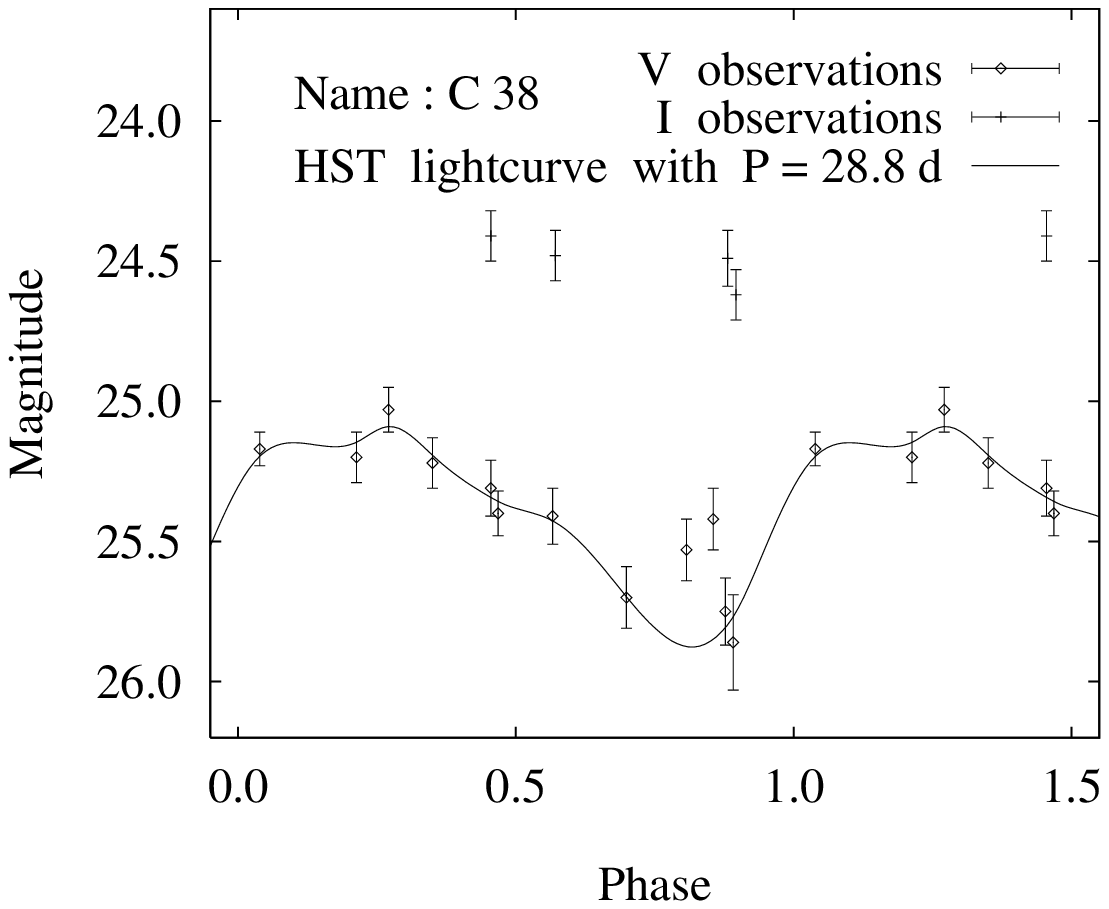}{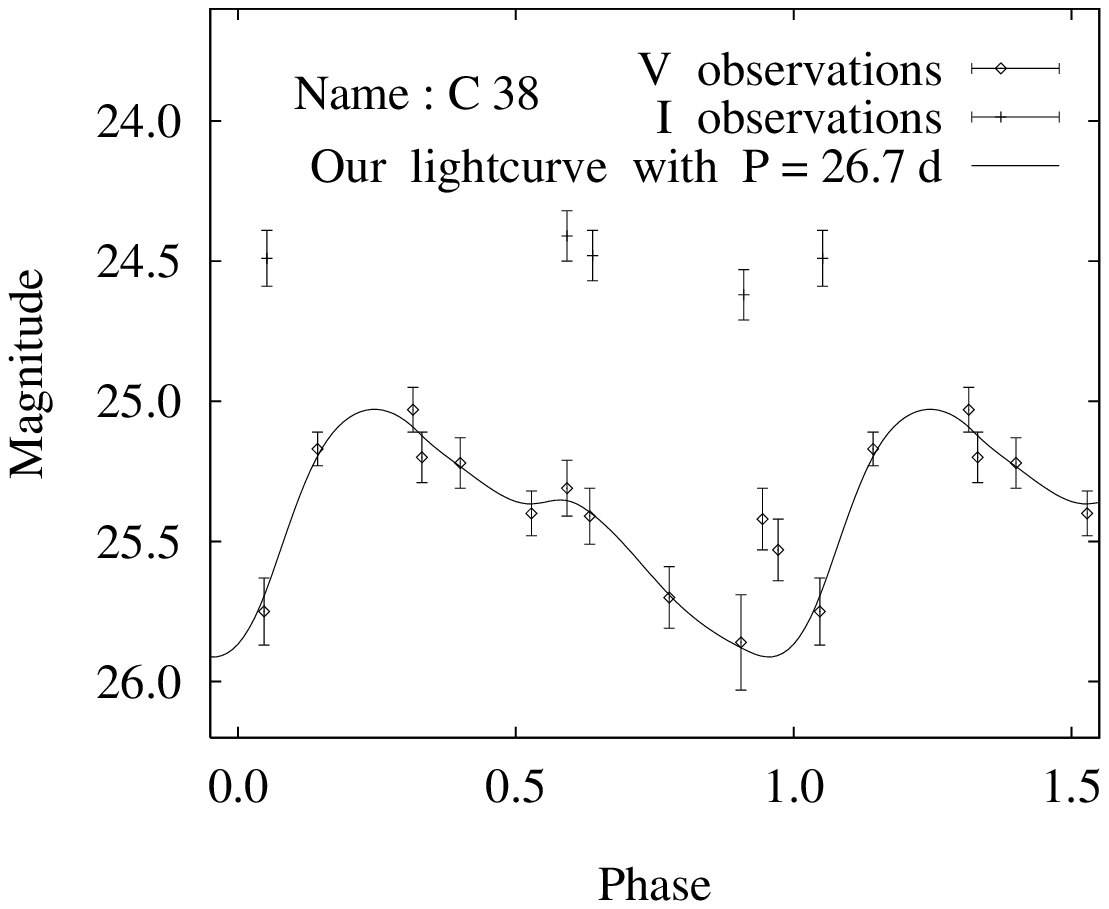}
{Light curves constructed for the same Cepheid with different periods
are compared. Our \lig\ with $P = 26 \fd 7$ resembling a characteristic
Cepheid \lig, has a better phase matching for the $I$ data than the
HST \lig\ with $P = 28 \fd 8$.}
{fig:ligc_comp}

From the plot of the number density of Cepheids against $\lp $
(Figure~\ref{fig:mw_m100}) we note that there is a dip in the number density
in the range $ 1.5 \leq \lp \leq 1.65$. In our view, this is caused owing 
to the lack of identification of a source as a Cepheid, a problem which becomes particularly severe while observing at fixed epochs
over a finite length of time which is comparable to the Cepheid period.
According to Saha and Hoessel (1990) in order to obtain a minimal \lig, it
is essential to have enough number of observations near both light maximum
and light minimum phases during a cycle. By observing at predetermined 
11 or 12 epochs one would invariably fail to detect the variation of light 
of a source within a certain range of periods at crucial points in its
\lig, and such a source could be easily missed as a variable source.
Indeed, such a dip is predicted from
the observing strategy by Ferrarese \ea (1996) also.

\figone{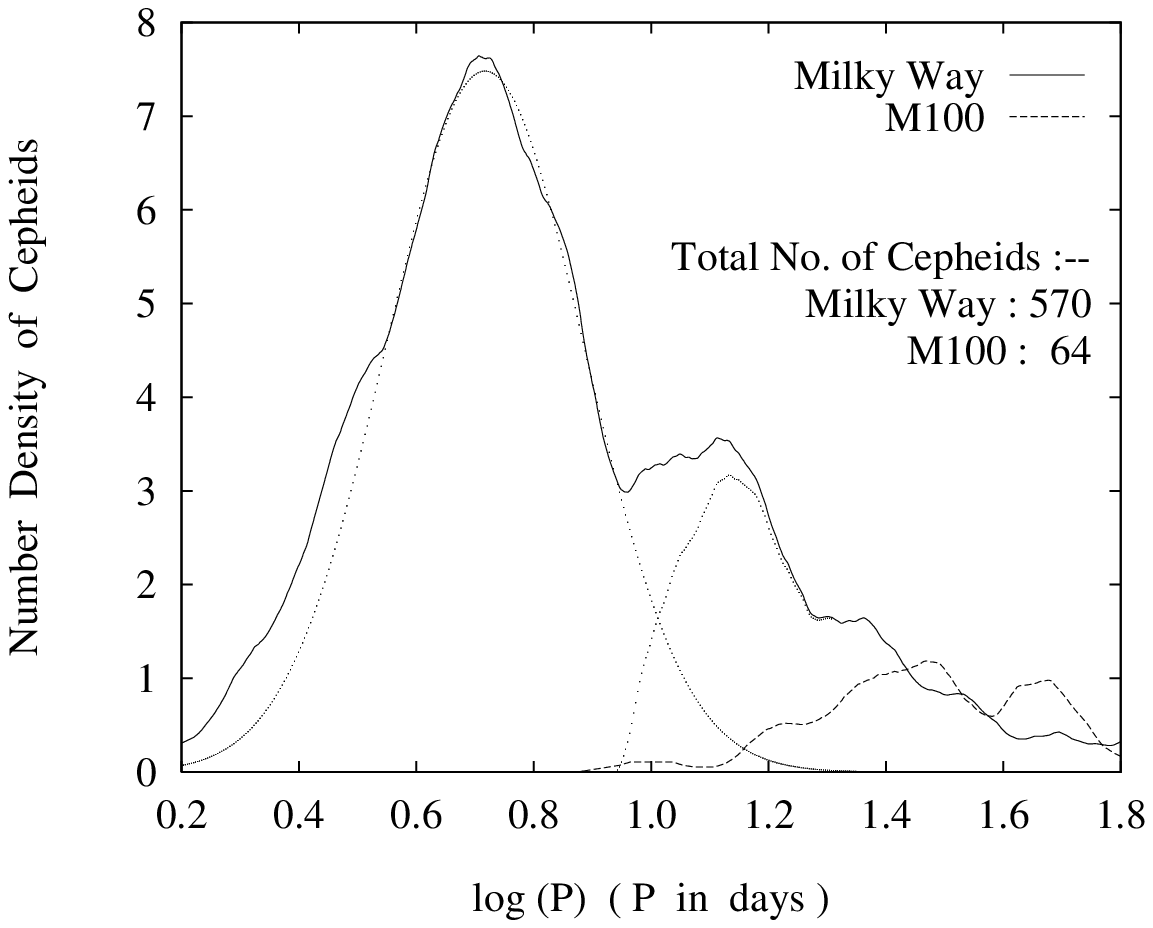}
{The number density distribution against $\lp $ is shown for Cepheids
in the Milky Way and M100. A moving average has been used to generate a
smooth curve from discrete observations.
The Galactic sample is split into two populations, having slight overlap
between periods of 9 and 18 days. The M100 sample lies almost fully in the
second population with $P \geq 9$ days.}
{fig:mw_m100}

We determined the V magnitudes by integration of the \lig s. We used
the method of synthetic \lig s (cf.\ Paper I) to obtain $\uncviav $.
Out of the 70 stars listed in Ferrarese \ea (1996),
ten have been excluded: three of them
due to their low period; two because we believe that their period is
much larger than the total time span of observation (56 days)
and consequently, the width of their plateau in the \lig\ near
minimum flux cannot be determined; four stars for which the \lig\
does not appear to conform to Cepheid variables for any of the converged
periods within the allowed range; one which had only one I band data.
Our final working sample consists of 60  Cepheid variables with a period 
range of 15 to 69 days. A comparison of our derived periods with those obtained 
by the HST Group is shown in Table~\ref{tab:cepparam} where we have
also given the mean $V$-magnitude ($\uncv $), $\uncviav $ color, 
amplitude of pulsation ($\vamp $) as well as the extinction-corrected 
$V$-magnitude ($\cv $) that we have estimated.

\begin{deluxetable}{ccccccccc}
\tablewidth{469pt}
\tablecaption{M100 Cepheid parameters
\label{tab:cepparam}
}
\small
\tablehead{
\colhead{} &
\multicolumn{2}{c}{Period (in days)}  &
\colhead{} &
\multicolumn{5}{c}{Photometric magnitudes from \lig s} \\
\colhead{\raisebox{2.4ex}[3ex][2ex]{ID }} &
\colhead{Ferrarese\tablenotemark{a}} &
\colhead{This work} &
\colhead{} &
\colhead{$\uncv $} &
\colhead{$\vamp $} &
\colhead{$\uncviav $} &
\colhead{$\uncvimax $} &
\colhead{$\cv $} }

\startdata
C1\ \ &  85.0 &\nodata& & \nodata & \nodata & \nodata & \nodata & \nodata \nl
C2\ \ &  66.2 &  69.0 & &  25.159 &   0.445 &   1.296 &   1.254 &  24.264 \nl
C3\ \ &  62.1 &\nodata& & \nodata & \nodata & \nodata & \nodata & \nodata \nl
C4\ \ &  54.0 &  55.2 & &  25.224 &   0.857 &   1.235 &   1.074 &  24.446 \nl
C5\ \ &  53.4 &  52.2 & &  25.035 &   0.655 &   1.075 &   0.947 &  24.640 \nl
C6\ \ &  52.0 &  45.9 & &  25.221 &   0.834 &   0.689 &   0.674 &  25.221 \nl
C7\ \ &  50.6 &  56.0 & &  24.758 &   0.671 &   0.891 &   0.868 &  24.758 \nl
C8\ \ &  50.2 &  51.0 & &  24.981 &   0.439 &   0.939 &   0.881 &  24.839 \nl
C9\ \ &  50.1 &  48.0 & &  25.886 &   0.597 &   0.989 &   0.947 &  25.522 \nl
C10 &  50.0 &  48.0 & &  24.702 &   0.674 &   0.899 &   0.766 &  24.702 \nl
C11 &  48.0 &  51.5 & &  25.492 &   0.828 &   1.127 &   0.913 &  24.953 \nl
C12 &  47.9 &  50.0 & &  25.304 &   0.778 &   1.323 &   1.063 &  24.435 \nl
C13 &  47.0 &  47.5 & &  25.483 &   0.612 &   1.121 &   0.911 &  24.963 \nl
C14 &  46.0 &  48.0 & &  24.945 &   0.806 &   1.023 &   0.854 &  24.667 \nl
C15 &  43.8 &  42.5 & &  25.438 &   0.853 &   1.075 &   0.864 &  24.999 \nl
C16 &  41.8 &  42.0 & &  24.908 &   0.634 &   1.086 &   0.929 &  24.471 \nl
C17 &  41.7 &  42.3 & &  24.982 &   0.804 &   1.139 &   1.138 &  24.982 \nl
C18 &  42.0 &  43.0 & &  25.211 &   0.958 &   0.937 &   0.743 &  25.050 \nl
C19 &  36.4 &  32.3 & &  25.749 &   0.787 &   0.901 &   0.616 &  25.749 \nl
C20 &  41.6 &\nodata& & \nodata & \nodata & \nodata & \nodata & \nodata \nl
C21 &  40.7 &  42.0 & &  25.324 &   0.787 &   0.948 &   0.889 &  25.068 \nl
C22 &  41.5 &  41.2 & &  25.439 &   0.687 &   0.589 &   0.540 &  25.439 \nl
C23 &  39.7 &  40.5 & &  25.640 &   0.657 &   1.101 &   0.904 &  25.106 \nl
C24 &  36.4 &  33.5 & &  25.670 &   0.915 &   0.843 &   0.504 &  25.670 \nl
C25 &  35.1 &  31.0 & &  25.853 &   0.392 &   0.628 &   0.510 &  25.853 \nl
C26 &  34.1 &  42.0 & &  25.885 &   0.779 &   1.253 &   1.123 &  24.997 \nl
C27 &  34.1 &  35.1 & &  26.151 &   1.137 &   1.036 &   0.764 &  25.709 \nl
C28 &  33.1 &  31.3 & &  25.611 &   0.353 &   0.834 &   0.833 &  25.611 \nl
C29 &  33.0 &\nodata& & \nodata & \nodata & \nodata & \nodata & \nodata \nl
C30 &  32.4 &\nodata& & \nodata & \nodata & \nodata & \nodata & \nodata \nl
C31 &  30.9 &  32.6 & &  25.499 &   0.672 &   0.741 &   0.695 &  25.499 \nl
C32 &  30.9 &  31.0 & &  26.067 &   1.197 &   0.760 &   0.304 &  26.067 \nl
C33 &  31.6 &  31.6 & &  25.766 &   1.412 &   0.961 &   0.743 &  25.378 \nl
C34 &  30.4 &  24.5 & &  26.132 &   1.308 &   0.802 &   0.343 &  26.132 \nl
C35 &  29.8 &  29.5 & &  26.171 &   1.090 &   1.286 &   1.087 &  25.061 \nl
C36 &  29.7 &  29.2 & &  25.417 &   0.675 &   0.618 &   0.487 &  25.417 
\tablebreak
C37 &  28.9 &  30.1 & &  26.180 &   1.015 &   1.219 &   1.123 &  25.236 \nl
C38 &  28.8 &  26.7 & &  25.395 &   0.721 &   0.915 &   0.744 &  25.346 \nl
C39 &  28.5 &  29.7 & &  26.162 &   1.026 &   1.190 &   0.902 &  25.371 \nl
C40 &  28.9 &  31.0 & &  26.069 &   0.786 &   0.996 &   0.966 &  25.559 \nl
C41 &  27.2 &  27.9 & &  24.896 &   0.633 &   0.784 &   0.746 &  24.896 \nl
C42 &  26.1 &  26.0 & &  25.855 &   1.069 &   0.976 &   0.767 &  25.494 \nl
C43 &  24.4 &  26.0 & &  25.539 &   0.928 &   0.520 &   0.468 &  25.539 \nl
C44 &  25.7 &\nodata& & \nodata & \nodata & \nodata & \nodata & \nodata \nl
C45 &  25.8 &  25.8 & &  25.672 &   1.080 &   1.008 &   0.820 &  25.223 \nl
C46 &  26.0 &  25.4 & &  25.336 &   0.966 &   0.548 &   0.332 &  25.336 \nl
C47 &  25.6 &  29.0 & &  26.153 &   0.711 &   1.055 &   0.885 &  25.604 \nl
C48 &  24.7 &  24.4 & &  25.961 &   1.071 &   1.029 &   0.682 &  25.671 \nl
C49 &  24.4 &  24.2 & &  26.300 &   1.243 &   1.093 &   0.749 &  25.683 \nl
C50 &  25.0 &  23.4 & &  26.182 &   1.476 &   1.069 &   0.551 &  25.570 \nl
C51 &  23.9 &  24.3 & &  25.976 &   1.201 &   1.009 &   0.683 &  25.676 \nl
C52 &  23.4 &  21.4 & &  26.575 &   0.949 &   0.804 &   0.458 &  26.575 \nl
C53 &  21.9 &  22.5 & &  26.521 &   1.216 &   0.941 &   0.816 &  26.050 \nl
C54 &  21.3 &  21.4 & &  26.219 &   1.281 &   1.088 &   0.530 &  25.548 \nl
C55 &  22.5 &  19.5 & &  26.474 &   0.851 &   1.118 &   0.823 &  25.869 \nl
C56 &  21.3 &  21.0 & &  26.250 &   1.858 &   0.766 &   0.372 &  26.250 \nl
C57 &  20.2 &  21.4 & &  26.436 &   0.560 &   1.246 &   1.065 &  25.592 \nl
C58 &  19.9 &  19.7 & &  25.733 &   1.050 &   0.587 &   0.447 &  25.733 \nl
C59 &  19.0 &  22.5 & &  25.491 &   0.960 &   0.619 &   0.555 &  25.491 \nl
C60 &  17.1 &  17.5 & &  26.174 &   0.855 &   1.184 &   1.027 &  25.337 \nl
C61 &  18.4 &\nodata& & \nodata & \nodata & \nodata & \nodata & \nodata \nl
C62 &  17.9 &  17.3 & &  26.100 &   1.374 &   0.961 &   0.567 &  25.788 \nl
C63 &  17.6 &  17.0 & &  26.092 &   1.213 &   1.193 &   0.843 &  26.092 \nl
C64 &  17.1 &  16.5 & &  25.803 &   0.787 &   0.510 &   0.484 &  25.803 \nl
C65 &  15.2 &  15.7 & &  25.937 &   0.729 &   0.529 &   0.357 &  25.937 \nl
C66 &  15.7 &  15.8 & &  26.280 &   1.121 &   0.451 &   0.154 &  26.280 \nl
C67 &  14.1 &  14.6 & &  26.458 &   0.774 &   1.315 &   1.308 &  26.458 \nl
C68 &  10.9 &  10.9 & & \nodata & \nodata & \nodata & \nodata & \nodata \nl
C69 & \ 9.2 & \ 9.2 & & \nodata & \nodata & \nodata & \nodata & \nodata \nl
C70 & \ 7.3 & \ 7.3 & & \nodata & \nodata & \nodata & \nodata & \nodata \nl
\enddata
\tablenotetext{a}{From Ferrarese \ea (1996)}
\end{deluxetable}
\clearpage

One major handicap of our analysis is the uncertainty in
the determination of $\uncviav $ in many cases; this is caused by the
extremely poor phase sampling in the I band.
In general, for reliable period determination, better phase coverage in at
least one band, effected by either larger number of observations, or better
choice of sampling phase points, is essential. As we shall see later, the
uncertainty in the period determination translates to an increased scatter
in the \pl\ diagram and leads to a substantial increase in the error
margin in the distance measurement.

\section{Period--Luminosity Diagram for the Classical Cepheids}
\label{sec:plr}

The slope and intercept of the \pl\ diagram is conventionally
derived by using the LMC Cepheids in the period range of 3 to 60 days and the
value of the slope is usually taken as $ -2.77$. However, as argued in Paper I,
at low periods,
many of the classical Cepheid variables are multi-mode pulsators and most often
the first or second overtone is the dominant mode of pulsation. But at
higher periods, by and large, the fundamental mode appears to be important. 
It has been established in Paper~I that we may split the parent Cepheid
population into two broad groups, according to their pulsation characteristics. 
For galaxies at far off distances, only the Cepheids with
high luminosity (i.e., those with higher periods, and probably with dominant
fundamental mode of pulsation) are detected. 
A comparison of the number densities of Cepheids at different periods
in the Milky Way and M100 (Figure~\ref{fig:mw_m100}) clearly shows the two
broad groups detected among Galactic Cepheids, signified by the two
peaks at periods around 7 and 16 days; on the other hand, in the M100
population, only the second group is detected, while the low-period Cepheids
are missed due to flux limitation.
The slope of the \plr\ derived by using all Cepheids is
heavily biased towards the low-period ones because of their numerical strength
in our neighborhood.
So a calibration of the \plr\ which is relevant for distant galaxies 
should be made in nearby galaxies {\em only with Cepheids of period
greater than 15 days}, avoiding contamination from low-period
pulsators, which have, on the average, a different slope in the \plr. 

We have calibrated the slope of the period--$V$ magnitude diagram by selecting
only the Cepheids in the period range of 15 to 70 days in LMC as well as
three spirals at moderate distances for which HST data is available.
To the extent possible from the available data, we have tried to do it with
extinction correction, as well as without; but since extinction correction
does not affect the slope of a complete sample, we have displayed the typical
numbers in Table~\ref{tab:slopes} without extinction correction.
It is readily seen from this table that there is a systematic 
change in the slope with increasing period ranges but within the range of 20
to 60 days it remains fairly constant. This effect was recognized earlier by
Morgan (1994) who had noted that a better fit to the \pl\ diagram is
obtained if the slope for higher period Cepheids is taken as $ -3.54$.
From a general study
of all these four galaxies, we arrive at an average slope of $-3.45$ for
Cepheids between periods of 15 and 70 days, with a possible error of 0.15.

\begin{table}[!h]
\small
\caption{
Slope of the Cepheid Period-Luminosity Relation 
from different galaxies
}
\label{tab:slopes}
\vskip 11pt
\begin{tabular}{lccccl}
\hline
\hline
\noalign{\vskip 8pt}
\multicolumn{1}{l}{Galaxy} &
{Period range} &
{No. of} &
{Slope of PL} &
{$\sqrt{\chi ^2/{\mathrm{d.o.f}}}$} &
\multicolumn{1}{l}{Source of the} \\
{} &
{} &
{Cepheids} &
{relation} &
{} &
\multicolumn{1}{l}{data}\\
\noalign{\vskip 4pt}
\hline
\noalign{\vskip 8pt}
LMC    &      all        &108     & $ -2.770 $ & 0.360 & Freedman (1995) \\
       & $\lp \geq 1.16$ &\phn 35 & $ -3.405 $ & 0.335 &                 \\
       & $\lp \geq 1.20$ &\phn 34 & $ -3.445 $ & 0.340 &                 \\
       & $\lp \geq 1.26$ &\phn 30 & $ -3.480 $ & 0.325 &                 \\
       & $\lp \geq 1.30$ &\phn 28 & $ -3.526 $ & 0.336 &                 \\
\tableline
NGC 925& $\lp \geq 1.27$ &\phn 30 & $ -3.615 $ & 0.294 & Silbermann \ea  \\
       & $\lp \geq 1.31$ &\phn 28 & $ -3.493 $ & 0.291 & (1996)           \\
       & $\lp \geq 1.35$ &\phn 24 & $ -3.447 $ & 0.297 &                 \\
\tableline
IC 4182& $\lp \geq 1.12$ &\phn 12 & $ -3.394 $ & 0.269 & Saha \ea (1994) \\
\tableline
NGC 3351& $\lp \geq 1.00$ &\phn 32 & $ -3.390 $ & 0.283 & Graham \ea (1997)\\
\tableline
\tableline
Our accepted & $\lp \geq 1.15$ &        & $ -3.450 $ &       &           \\
value        &                 &        & $\pm 0.15$ &       &           \\
\noalign{\vskip 5pt}
\hline
\end{tabular}
\end{table}

Just as the slope of the period--$V$-magnitude relation needs
to be determined for the representative sample of Cepheid variables,
it is equally important to find the scatter in the relation
if we intend to provide a trustworthy error analysis of our
distance estimations.  The data we have used cannot  provide
a good estimate  for the  scatter  in the extinction corrected 
period--$V$-magnitude relation for Cepheids
of period in the range of 15 to 70 days which are in their second passage
of the instability strip. A value of 0.20 to 0.25 magnitude appears
to be indicated by our analysis of the HST data for some of the galaxies,
but in the absence of a robust method of positioning them in the
\gteff\ plane, the estimation for the intrinsic scatter should be
taken with caution.

The zero point of the Cepheid \plr\ is another issue for extensive debate.
Conventionally, it is calibrated by assuming a distance modulus to LMC.
However, since the quoted distance modulus to LMC ranges from less than
18.35 to more than 18.7 mag, we preferred to isolate the \plr\ from this 
secondary calibration. Recent trigonometric parallax observations by the 
Hipparcos satellite (\cite{fc:97}) provides us with direct distances to 
some of the nearby Cepheid variables. We have used the zero point calibration
of this work and adopted the Hipparcos value of $ -4.24$ for the mean
absolute $V$-magnitude for a Cepheid of period 10 days. The same result
is arrived at if we take only the three Cepheids of period $> 10$ days from
their sample. Hence we have arrived
at the following \plr\ for Cepheid variables of period greater than 15 days:
\beq
M_V = -3.45\, \lp - 0.79
\label{eq:plr}
\enq
We assign a zero point error of 0.20 magnitude, though there is continuing 
debate on (a) whether Hipparcos parallax data systematically underestimates
the distance to stars, and (b) whether the Feast and Catchpole (1997)
calibration provides only an upper limit to the Cepheid distance scale.

\section{Incompleteness Correction}
\label{sec:incomp}

A major task in the extragalactic distance measurement, like the
HST observations of Cepheid variables in M100,
is to isolate the signal from the noise near the limiting magnitude
at which a precise determination of the stellar magnitude is barely feasible.
We argued in Section~\ref{sec:per_det} that the typical signal to noise 
ratio (SNR) for the HST observations of stars in M100 with filter F555W is 
around 6--8 for stars of $V$-magnitude near 26 and of the order of 10--15
for $V$-magnitude near 24.5. This rapid change in the SNR causes faint
stars to be systematically missed in the sample, while the brighter
stars preferentially detected at a fixed period produce an increase in
the average brightness of the stars at that period, if the scatter in the
period--$V$-magnitude diagram is large. (For instance, the HST data
before extinction correction has scatter in the $V$-magnitude of order 
0.45 magnitude at a fixed period). We now address the question:
with this scatter in $V$-magnitude, is there a systematic over-estimation
of the brightness of the M100 Cepheids at low periods? (i.e., is there the
effect known as Malmquist bias (cf.\ \cite{sandage:87}) in this sample?)

\figone{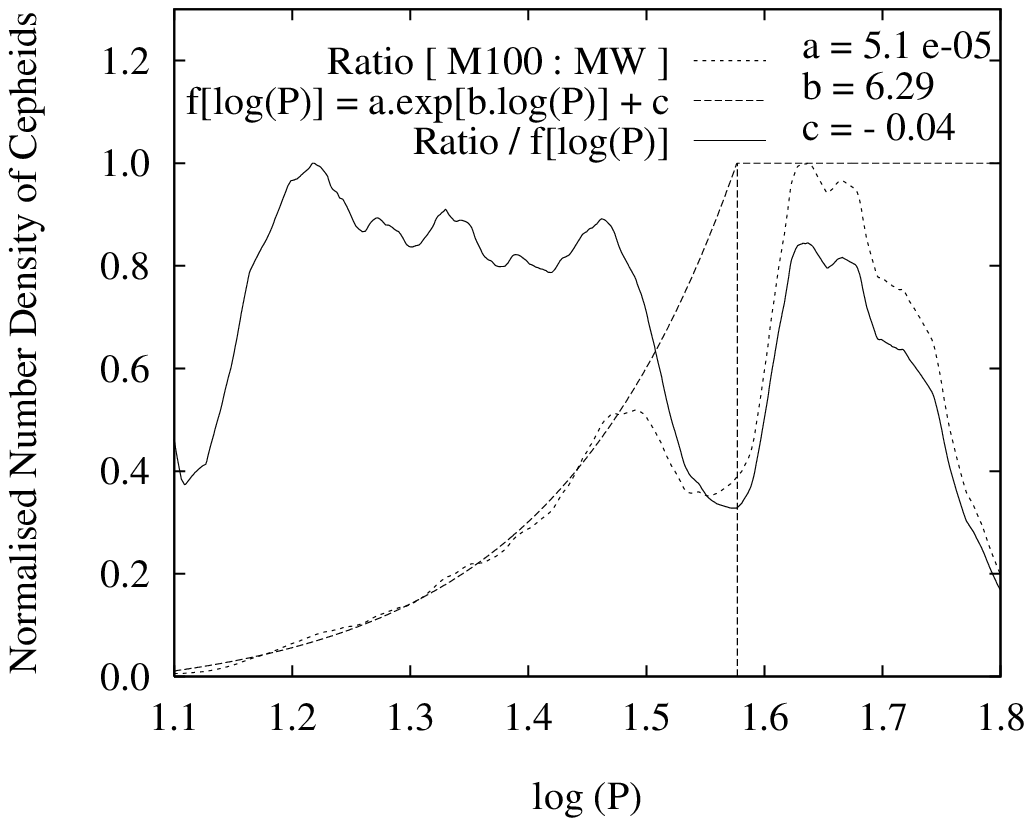}
{The ratio of number density distribution of Cepheids in M100 and
the Galaxy is displayed (dotted curve). At low periods, an exponential
curve [$5.1 \times 10^{-5} e^{\{ 6.29\, \lp \} } - 0.04$] is
seen to fit this ratio well; when scaled by this function, the curve is
seen to be free from flux-dependent incompleteness (solid curve).}
{fig:m100_ratio}

We can compare the HST sample
of Cepheid variables in M100 with the reasonably complete sample of Galactic
Cepheids (discussed in Paper I) to estimate this systematic effect.
In Figure~\ref{fig:mw_m100} we have displayed the observed number as a function
of their period after carrying out a moving average for both the Galactic
and M100 Cepheid variables.
It is evident that almost all the HST data pertains to the Cepheids of period
greater than 10 days, which lie in the second component of the Galactic
distribution function (see Paper~I). The ratio of the
number density of the M100 Cepheids to that of the Milky Way is plotted
as a function of the period in Figure~\ref{fig:m100_ratio},
where we note the following features:
\bei
\item
The ratio tends to a constant at large periods which is not surprising
since the HST data has no
systematic incompleteness at higher periods and also the galaxy M100 is
similar to the Milky Way.  But due to small number statistics of the 
Galactic sample, the shape of the graph
is not reliable beyond a period around 50 days.
\item
At intermediate periods, there is a dip in the ratio indicating that
most of the Cepheids at period of around 40 days have probably been missed.
We argue later that this is owing to the fact that
the HST data spans a duration of 57 days and observations were
carried out essentially at 11 epochs.
\item
At period lower than 30 days, the ratio between the number densities
of M100 and Galactic Cepheids indeed shows a sharp decrease which we 
attribute to
a systematic incompleteness due to the sample being flux--limited.
The fall in the ratio with decreasing frequency is
almost exponential, suggesting that the detection efficiency probably varies
as some power of the number of photons received.
\eni
However, as we have demonstrated in Paper~I, from the HST observations of 
Cepheids in galaxies at closer distances, the pattern of the HST distribution
function appears to be similar to that of Galactic Cepheids, indicating that
at magnitudes brighter than 22 mag, the effect is negligible.

The extent of
incompleteness in the sample of M100 Cepheids is a combined effect
of $(i)$ magnitude-limited detection efficiency at low periods,
$(ii)$ intrinsic scatter in the \plr, and
$(iii)$ increased scatter in the observed period--$V$-magnitude diagram
due to errors in extinction correction as well as uncertainty in the
determination of the pulsation period.

The detection of a signal as function of the number of data points and
the signal to noise ratio was discussed in Section~\ref{sec:per_det}.
If $N_{\mathrm{ph}}$ is the number of photons arriving at the telescope 
from a star,
then it can be easily derived that the signal-to-noise ratio for bright
objects is proportional to $\sqrt{N_{\mathrm{ph}}}$, while for dim objects, 
it is proportional to $N_{\mathrm{ph}}$ itself.
Thus, for sufficiently bright stars the
probability of detection as a Cepheid with the HST scheme for M100 is
practically a constant and the sample can be treated as though it was
volume-limited. But for low period Cepheids the detection of pulsation 
with the 11 sample points becomes inefficient, although the 
observed efficiency would naturally depend on the algorithm used. Instead of
going through these details, we can represent the detection efficiency
for stars fainter than some cutoff magnitude $V_{0}$ by
\beq
D(V) \sim \exp [-\gamma (V - V_{0})/\alpha ],
\enq
where $\gamma $ is a constant determined by the detector characteristics, 
$\alpha $ is the magnitude of the slope of the period--$V$-magnitude diagram,
and $V$ is the mean apparent magnitude of the star.
The consequences of this for the distribution function of the Cepheids
at a fixed period are discussed in the Appendix.
At $V \gg V_{0}$, the incompleteness correction tends to a constant
that depends only on the scatter in the period--$V$-magnitude diagram
and the relation between SNR and efficiency of detection; this is
because the stars mainly at the brighter end of the intrinsic distribution
function at a fixed period are detected. At $V \ll V_{0}$, the correction
tends to zero as to be expected.

As already noted in Section~\ref{sec:per_det}, the
number density profile of M100 Cepheids shows a dip in the range $ 1.5 \leq
\log (P) \leq 1.65 $ caused due to the failure of detection of Cepheids.
However, this decreased efficiency has no direct relevance to
flux-limited incompleteness and we do not analyze it further.
However, as we have discussed in the Appendix, the dip does introduce a
systematic error when we try to determine the magnitude at which 
flux-limited incompleteness can be ignored, and consequently does affect
the distance calibration.

If the probability density of the Cepheids as a function of the intrinsic 
magnitude retains its form and has the same scatter ($\sigma $) when
the pulsation period
varies, then any error in the determination of the period only increases
the scatter in the observed period vs $V$-magnitude diagram.
The error in the period can be incorporated in the observed
probability density by replacing $\sigma $ by an effective scatter
($\sigma_{\mathrm{eff}}$) in the observed distribution, given by
\beq 
\sigma_{\mathrm{eff}} = [ \sigma_{\mathrm{int}}^2 + \frac{1}{6} \alpha^2
\sigma_{\mathrm{P}}^2 ]^{1/2} 
\enq
correct to second order. Here $\sigma_{\mathrm{int}}$ is the intrinsic
scatter in the \pl\ diagram, while $\sigma_{\mathrm{P}}$ is the uncertainty
in $\lp $. Based on our period determination methods,
the value of $\sigma_{\mathrm{P}}$ is estimated to be around 0.1. For
Cepheids with periods less than 30 days it is usually less than this value,
while for $P > 30$ days, it lies between 0.1 and 0.2. For periods higher
than 50 days, it is difficult to estimate the value of $\sigma_{\mathrm{P}}$. 
If the extinction correction can be carried out in a manner independent of the
incompleteness, $\sigma_{\mathrm{eff}}$ is the scatter in the 
period--$V$-magnitude
diagram after the correction is applied. But in our scheme, we cannot
carry out the correction independently of incompleteness as we use
the period vs color, $V$-amplitude vs color-at-peak-brightness and the
period--$V$-magnitude relations simultaneously. Consequently, the
$\sigma_{\mathrm{eff}}$ we used is intermediate between the values in 
Figures~\ref{fig:incomp_comp} and~\ref{fig:v_logp}. 

We have carried out the correction for incompleteness bias
as discussed in the Appendix, and our prescription to obtain a complete
sample is
\beq
V_{\mathrm{complete}} =
  \cases{
     V_{\mathrm{incomplete}} \,+\, \sigma_{\mathrm{eff}}^2 \,
     \frac{\gamma}{\alpha} 
     & {for $\lp \leq 1.52$}\cr
     V_{\mathrm{incomplete}} \,+\, \sigma_{\mathrm{eff}}^2 \,
     \frac{\gamma}{\alpha} \,
     \frac{1.64 - \lp}{1.64 - 1.52}
     & {for $1.52 < \lp \leq 1.64$}\cr
     V_{\mathrm{incomplete}}
     & {for $\lp > 1.64$}
 } 
\label{eq:incomp}
\enq
Within the observational errors this prescription agrees with the more
detailed numerical results given in Table~\ref{tab:incomp} based on the
formulation described in the Appendix.

\begin{table}[!h]
\small
\caption{Incompleteness Correction based on Numerical Simulations
\label{tab:incomp}
}
\vskip 11pt
\begin{tabular}{ccccccc}
\hline
\hline
\noalign{\vskip 8pt}
\multicolumn{1}{l}{Cutoff} &
\multicolumn{1}{l}{Effective} &
\multicolumn{5}{c}{$\delta V$ for period $\lp $} \\
\multicolumn{1}{l}{Magnitude ($V_{0}$)} &
\multicolumn{1}{l}{Scatter ($\sigma_{\mathrm{eff}}$)} &
{$< 1.40 $} &
{$ 1.50 $} &
{$ 1.55 $} &
{$ 1.60 $} &
{$ 1.70 $}\\
\noalign{\vskip 4pt}
\hline
\noalign{\vskip 8pt}
  25.0  &       0.36   &        0.26 &   0.25 &  0.23 & 0.22 & 0.17 \\
  25.4  &       0.36   &        0.24 &   0.20 &  0.18 & 0.15 & 0.11 \\
  25.6  &       0.36   &        0.21 &   0.17 &  0.15 & 0.12 & 0.09 \\
\hline
  25.0  &       0.42   &        0.35 &   0.33 &  0.30 & 0.29 & 0.23 \\
  25.4  &       0.42   &        0.31 &   0.27 &  0.24 & 0.20 & 0.16 \\
  25.6  &       0.42   &        0.29 &   0.23 &  0.19 & 0.16 & 0.13 \\
\hline
\multicolumn{2}{l}{Value used in this work} & &       &      &   &  \\
\multicolumn{2}{l}{(from Figure~\ref{fig:m100_ratio})}&
                                 0.28 &  0.28  &  0.21 & 0.09 & 0.00 \\
\hline
\multicolumn{2}{l}{Mean difference} & 0.00 & -0.04 & 0.01 & 0.10 & 0.15 \\
\noalign{\vskip 5pt}
\hline
\end{tabular}
\end{table}

We should stress that this is a statistical method, where instead of
increasing the mean magnitude at a fixed period by the specified correction
term, we increase the magnitude of each star in that period range. The 
incompleteness-corrected $V$-magnitudes are shown along with the observed
magnitudes in Figure~\ref{fig:incomp_comp}. Note however, that
the extinction-corrected $\cv $ values in Table~\ref{tab:cepparam} are the
{\em true} magnitudes for each star, not the incompleteness-corrected 
expectation value at the particular period.

\figone{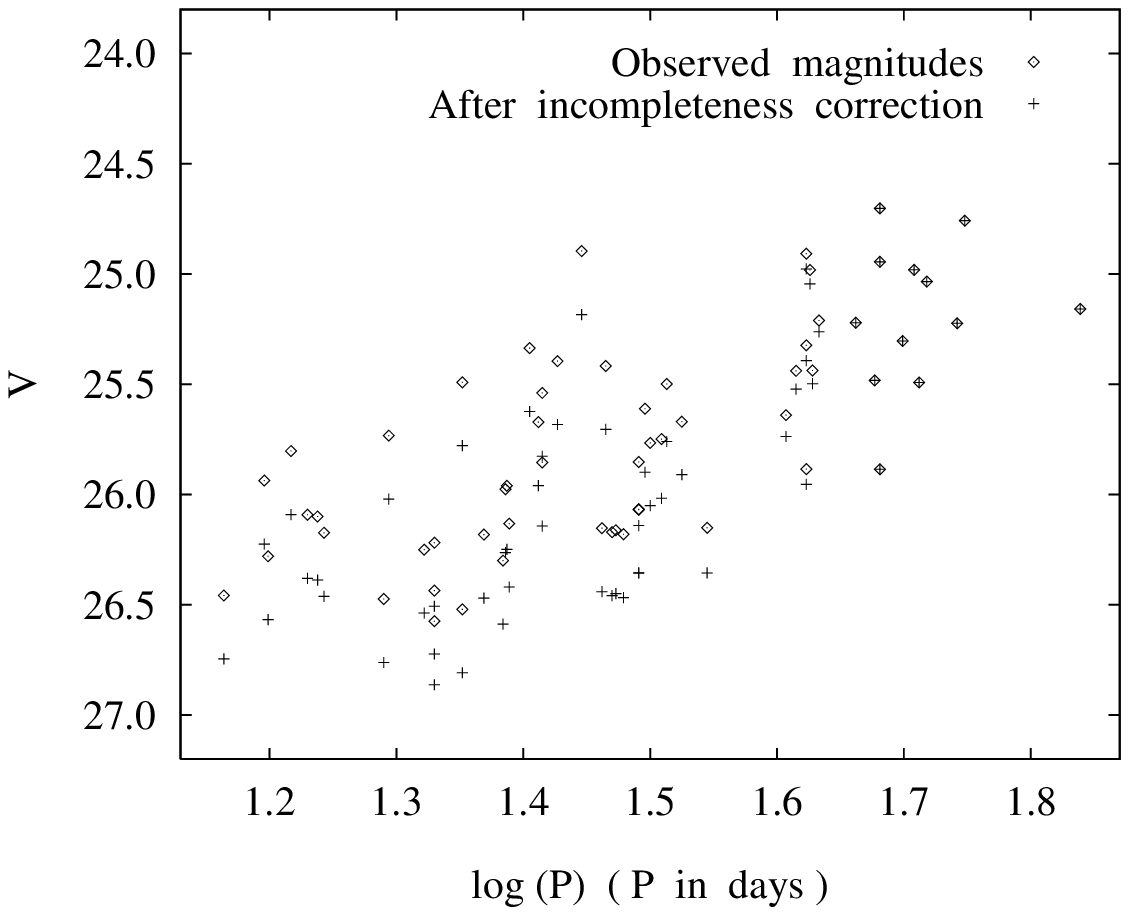}
{The raw $V$ magnitudes of M100 Cepheids (obtained by integrating the
\lig s) and the statistically incompleteness-corrected magnitudes (according to
Equation~\ref{eq:incomp}) are plotted against $\lp $.}
{fig:incomp_comp}

\section{Extinction Correction}
\label{sec:extcor}

Extinction correction is important for distance calibration even for
a face-on spiral like M100, since the stars observed at low periods
would be predominantly of low extinction while at higher period the
mean extinction is expected to be larger.
In Paper~I, we derived a formalism for the reddening and extinction correction
for Galactic Cepheids, based on their multi-wavelength observations.
However, in the absence of
multi-color photometry or at least full \lig s in two bands, the
extinction correction carried out would be at most statistical in nature
and would not take into account the differential extinction with respect
to period. For want of better alternatives we have used the three relations,
namely, $\viav $ vs $\lp $, $\vimax $ vs $\vamp $ and
$\cv$ vs $\lp $, for distance calibration as well as extinction correction.
However, we did not use pre-determined slope or intercept for any of these 
relations; instead, we resorted to $L_1$ minimization to obtain these six 
unknown quantities.

We prefer to use $L_1$ minimization over the standard practice of $L_2$
($\chisq $) minimization for the following reason.
The M100 data has large observational error bars, and since neither can 
we estimate the extinction well enough, nor can we identify the Cepheids
in a different evolutionary stage (cf.\ Paper I), the scatter in the \pl\
diagram as well as in the \pca\ relations remains large. The $L_2$ minimization 
is much more sensitive to such noise in the data where the
points having large error bars affect the mean values appreciably.
Barrodale and Zala (1986) argue that
$L_1$ is far more accurate in linear programming like ours where a few
points far from the straight line should not contribute substantially 
so as to change the slope or the intercept. The $L_1$
minimization leads to a value which is closer to the median of the sample,
rather than the mean that would be expected from $L_2$. Also, they argue
that ``trimming'' is 
advisable for $L_1$, which we have used only to distinguish between
similar minima when we try a range of slopes and intercepts in the
six-dimensional parameter space involving period vs $\cv $, period vs $\viav$
color and amplitude vs $\vimax $ minimization.

We have adopted the linear relation between the reddening-corrected 
$\viav$ and $\lp$,
\mbox{$\vimax$} and $\vamp$ as well as extinction and incompleteness bias
corrected $\cv $ and $\lp$ for $L_1$ minimization.
We minimize the absolute deviation $\chi_1$, defined by
\beqar
\chi_{1} & = & 
\sum_{i} a_1 | \langle V \rangle_{0i} + \alpha \log (P_i) - \mu | 
    + a_2 | \langle V-I \rangle _{0i} - \beta_1 \log(P_i) - y_1 | \nonumber \\
 &  &  \mbox{} + a_3 | \{ \vimax \} _{i} + \beta_2 \Delta V_i - y_2| 
\enqar
Ideally the weights $a_1$, $a_2$ and $a_3$ should be determined from the
error estimates in the photometry as well as from the scatter in the three
relations. We have chosen the three weights to be 0.2, 0.5 and 0.3 respectively 
in order that the scatter in both the best fit lines for $\vi $ are 
comparable to the errors in the observed colors in our data.
The deviation $\chi_1$ can be computed for a specified set of
parameters $\alpha$, $\beta_1$, $\beta_2$, $\mu$, $y_1$ and $y_2$
by choosing the reddening $\evi $ and extinction $\av $ for each star.
We have chosen a constant ratio $\av / \evi = 2.44$ in the absence of
multi-band observations (see e.g., \cite{ccm:89}). The mean dereddened 
color and its value at the brightest phase are both taken to be
$\viav = \uncviav - \evi$ and $\vimax = \uncvimax - \evi$.

For each data point the minimum deviation will be produced either at
zero extinction or when the point falls on one of the three straight lines,
subject to $\evi >0$. The minimization of $\chi_1$ with respect
to the intercept $\mu$ in the period--$V$-magnitude relation
provides the estimate for the distance modulus if the other parameters
are fixed.
The respective slopes  $\beta_1$ and $\beta_2$ for the Galactic Cepheids
were found to be 0.13 and 0.28 (cf. Paper I) and in 
Section~\ref{sec:plr} we argued that $\alpha \sim 3.45$.
But since the statistics for the Galactic sample is not very robust,
we kept the two intercepts for the color diagram as well as all the
three slopes to be unknowns having narrow
range of acceptable values.  For the period--$V$-magnitude relation we scanned
for slopes between $-3.30$ to $-3.60$ to obtain a value of $-3.49$, though for
higher incompleteness correction at certain periods a slope of
$-3.52$ and a higher intercept would be preferred.
Similarly, our $\chi_1$-minimized values  $\beta_1 = 0.13$ and
$\beta_2 = 0.30$ match with the Galactic values within 0.02. However,
the intercepts, $y_1 = 0.69$ and $y_2 = 0.94$ show a difference of
0.02 and 0.05 respectively, but in view of the corrections not included 
(which we discuss below), we consider the intercepts to be consistent with
their Galactic counterparts. This agreement in spite of substantial error
in the individual $\uncviav$ values makes us trust the final distance
modulus to at least within the errors we have given. The average extinction,
$A_V$ is 0.30 mag which agrees with the results of other workers but in 
view of the Malmquist bias of 0.28 mag at low periods, a value of 
$\langle A_V \rangle \approx 0.20$--$0.25$ would have been comfortable.

It should be noted that many of the points lie exactly on the lines in the 
three plots (Figures~\ref{fig:v_logp} and~\ref{fig:virel}); this is a
natural consequence of the $L_1$ minimization where the expectation value
is closer to the median than to the mean. We shall like to again stress that 
the procedure automatically assigns less weightage to the few data points far
from the line either due to large errors or due to the star being at a
different stage of evolution.

The following effects could not be studied quantitatively and hence
their contribution to the error in the distance modulus cannot be
ascertained: 

The four fields of M100 where Cepheids were observed will contain numerous
unresolved red dwarfs. Their presence is unlikely to change the $V$-magnitude
of the Cepheids but they could modify the $\vi $ \lig, though the
variation will depend upon
the method employed to subtract the background. This will affect the extinction
correction as well as possible tests on metallicity of the Cepheids in
M100, but in the absence of reliable I band \lig\
we cannot carry out quantitative analysis of this effect.

The Galactic Cepheids  which are not at the
second passage of the instability strip follow different \pca\ relations.
It was shown in Paper~I that their presence in the sample can increase
the slope of the $\BVav $ vs $\lp $ relation from $\sim 0.2$ to nearly 0.4 and
that they are detectable from their conspicuously different positions on 
the \gteff\ plane. However, for the M100 Cepheids with
only sparsely sampled observations
in two bands, neither the position of the instability strip on the \gteff\
plane can be determined, 
nor can the pulsators at different evolutionary stages be identified.
The  contamination from stars at first or third passage of the
strip will introduce errors in the final \pca\ and \pl\ relations of M100
Cepheids.

\figone{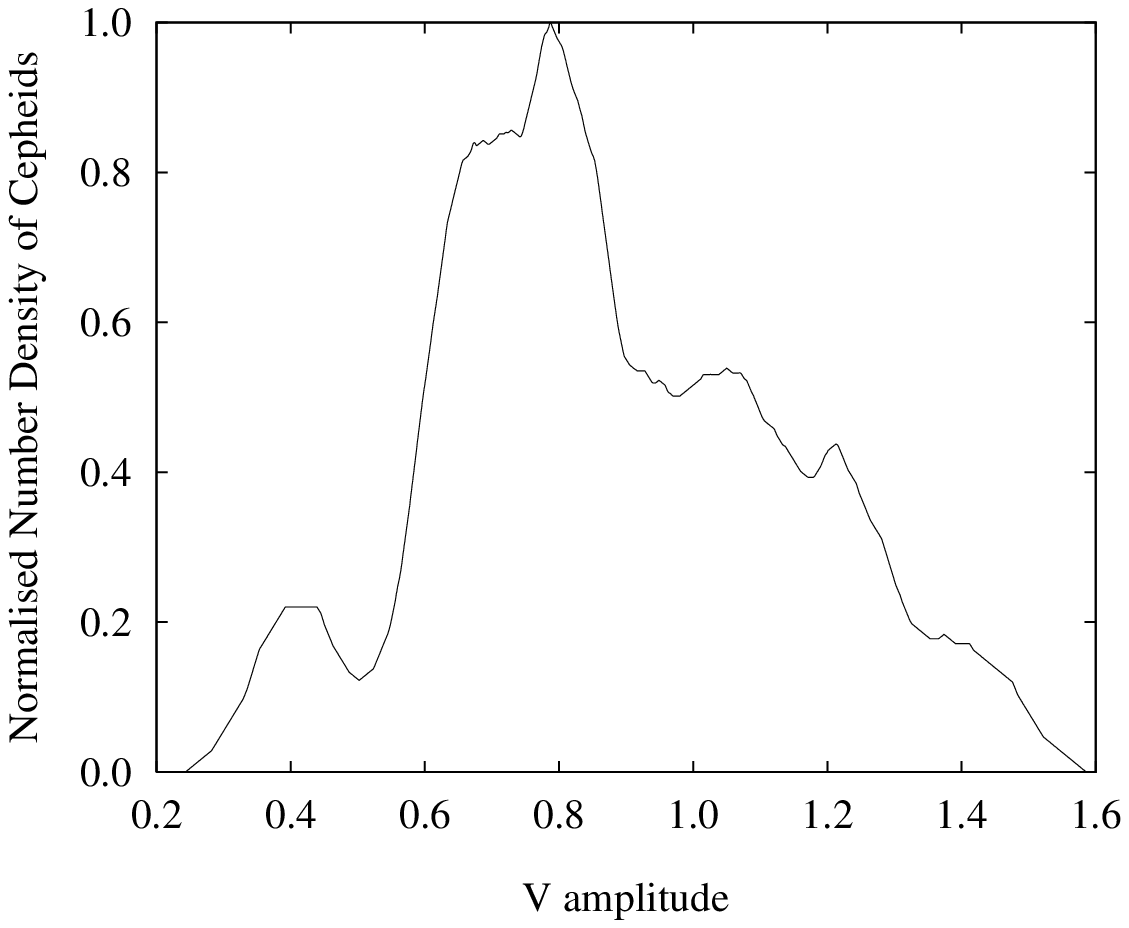}
{The normalized number density of M100 Cepheids as a function
of their $V$ amplitudes of pulsation are shown. A moving average has been
used to generate a smooth curve from discrete observations.}
{fig:ampl_dist}

The Galactic Cepheids of period larger than 15 days have an average 
V band pulsation amplitude of $\sim 1.1$ mag and most of these
variables have amplitude greater than 0.8 mag.  
But from Figure~\ref{fig:ampl_dist},
it is evident that the M100 Cepheids in the same range of period have
$V$-amplitudes  generally lower than their
Galactic counterparts, the reason for which is not known.
Moreover, the amplitude appears to decrease on the average, when
the period of pulsation increases. Both these effects could be
an artifact of the sparse sampling or due to observations of
specific regions in M100, unlike  the Galactic sample which is not confined
to any part of the Milky Way. Nevertheless, the amplitude vs
color-at-maximum relation should be scrutinized to examine whether
we compare similar samples in two different galaxies.

\section{Results and Discussion}
\label{sec:results}

The final result for the Cepheid variables in the spiral M100 in Virgo
Cluster, after corrections for extinction and incompleteness of the sample
and after carrying out the $L_1$ minimization, is given by the 
period--$V$-magnitude relation (Figure~\ref{fig:v_logp})
\beq
\cv = -3.49 \lp + 30.80,
\label{eq:v_logp}
\enq
the period--color relation (Figure~\ref{fig:virel})
\beq
\viav = 0.13 \lp + 0.69,
\enq
and the $V$-amplitude--color-at-brightest-phase relation 
(Figure~\ref{fig:virel})
\beq
\vimax = -0.30 \vamp + 0.94
\enq

\figone{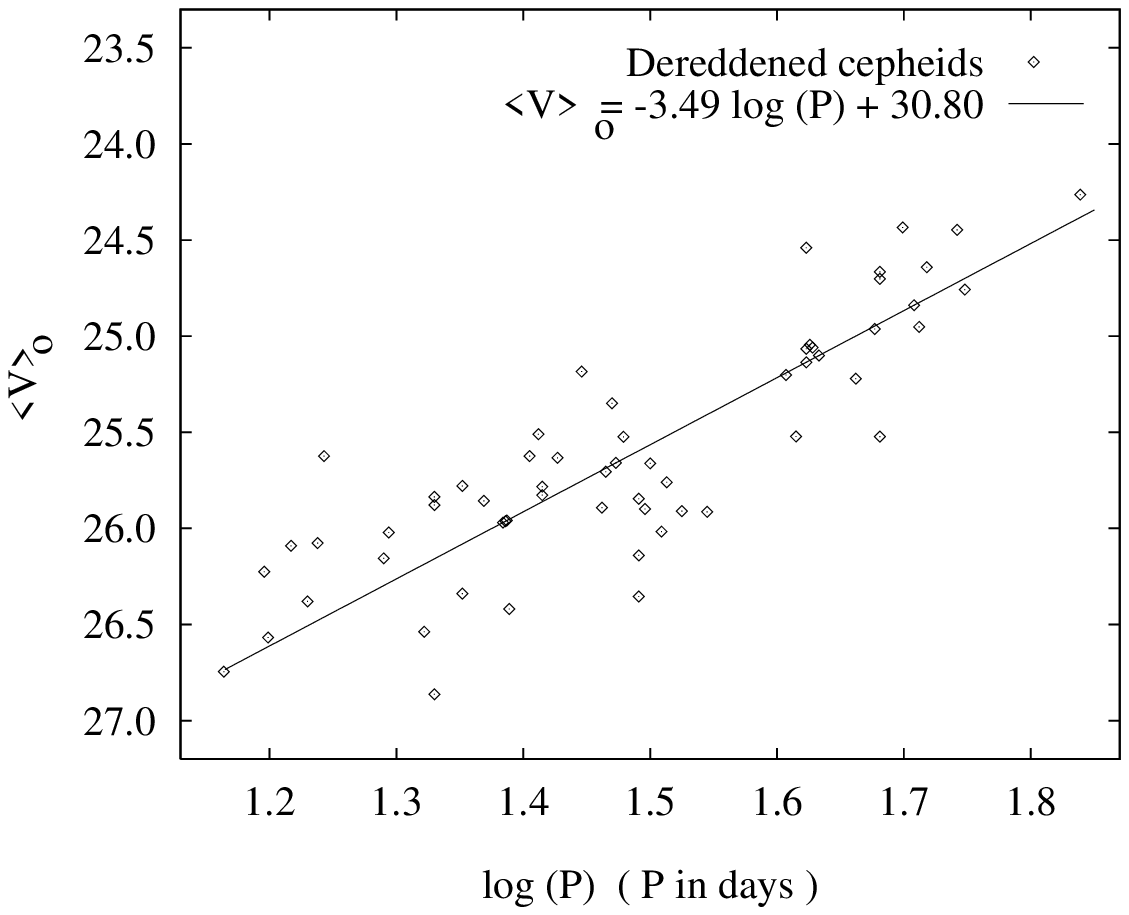}
{The final PL diagram for M100 Cepheids is shown, along with the plot
for the best fit \plr.}
{fig:v_logp}

\figtwo{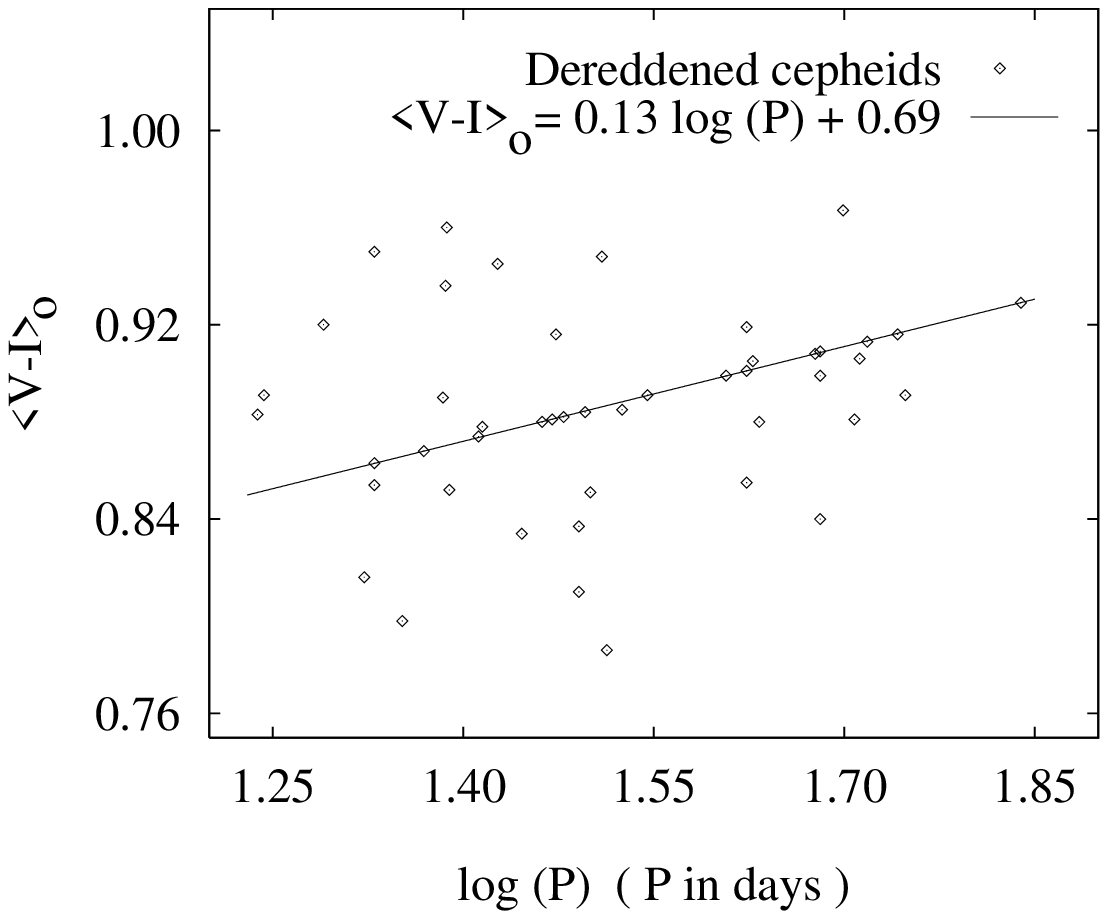}{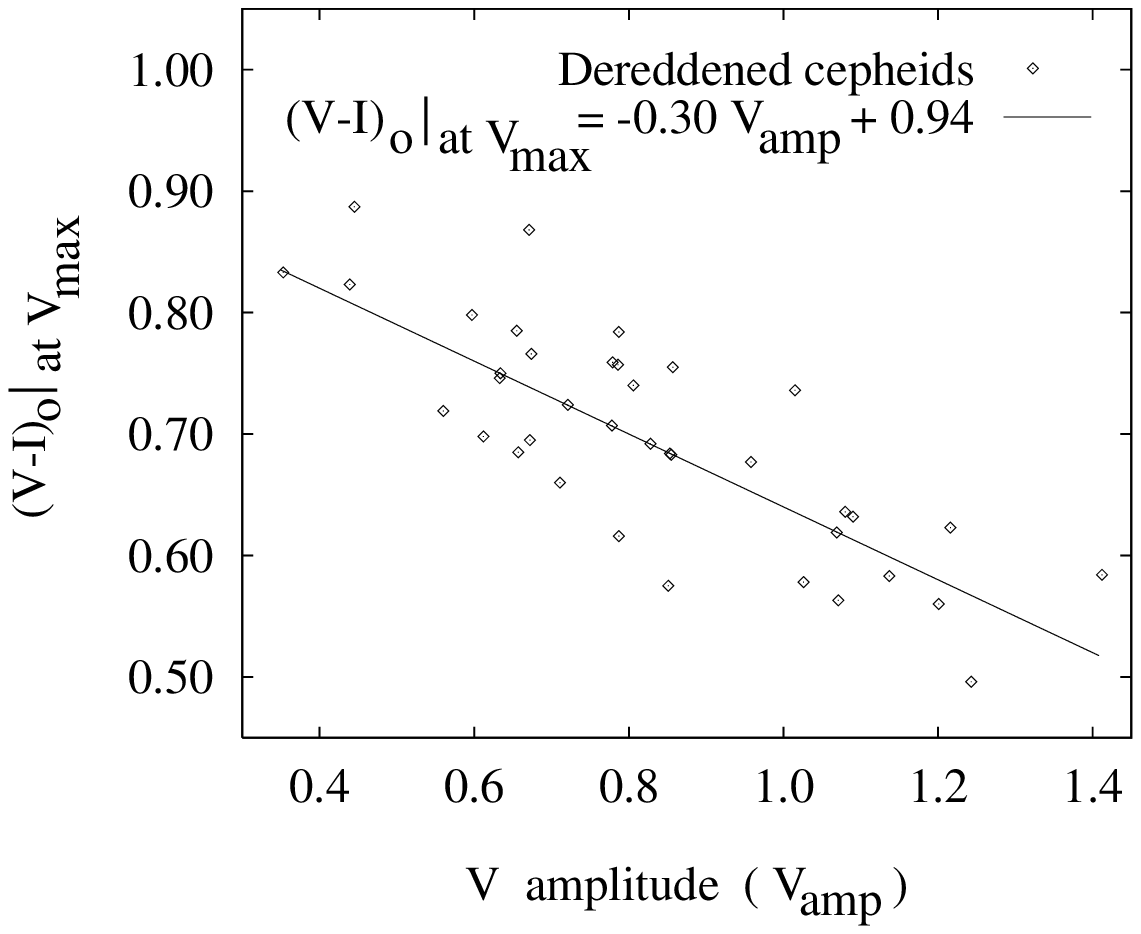}
{Period--color and amplitude--color plots for the
M100 Cepheids are shown along with the best fit linear relations.}
{fig:virel}

If the \pca --luminosity relation for the Galactic
Cepheids had been better established and the I band \lig\ of the
Cepheids in M100 were observationally determined, the minimization would have
been possible with fewer unknown parameters. This would have provided better 
tests of the result as well as of the internal consistency of the
data. The specific issue is: does the sample observed in M100 belong to the
same population to which the Galactic Cepheids belong to, or are the
Galactic Cepheids studied so far subject to systematic biases?
Similarly, if the errors in the period determination were fewer, there
would have been less mixing in the observed period--$V$-magnitude diagram
and the scatter in the diagram could have been considerably less
at periods in the range of 35 to 45 days, which is the
crucial region in determining the distance modulus.
At present, the error estimates should be treated as indicative only, since
we do not have sufficient internal checks on them.

Here we discuss some of the additional problems which,
we feel, should be addressed adequately in future key projects:

The  mean extinction correction is only as reliable as the colors even if the
period is known accurately. We estimate that the $\vi$ data has approximately
0.15 magnitude error. Since the average reddening is of the order of
0.10 and probably much less for the variables of low period,
many of the Cepheids show negative reddening. In our treatment, these
stars would have zero extinction and this systematically overestimates
the mean extinction. When the Malmquist bias for Cepheids fainter than 25.2
magnitude is taken into account for the present M100 sample, we find that
half the Cepheids are susceptible to this error
at period of less than 40 days. This was also borne out by our analysis.
Should our estimation of the cutoff magnitude for flux-limited bias
of 25.2  magnitude be correct, we might be overestimating the
extinction by approximately 0.1 magnitude and ultimately the distance to
M100 by the same amount.

In Figure~\ref{fig:m100_ratio} we were unable to determine the shape of 
the curve (relative number density
of Cepheids in M100 to that in the Milky Way, as function of the
pulsation period) at 35 to 45 days because of the dip in the 
number density of Cepheids detected in M100 due to extraneous reasons.
This has indirect implications for the correction to the flux-limited
detection efficiency as is discussed in the Appendix and it
could possibly explain some skewness in Figure~\ref{fig:v_logp}.

Within these limitations, the resulting period--$V$-magnitude relation
for the Cepheids in M100 can be compared with the one we have arrived at
in Section~\ref{sec:plr} to determine the distance modulus to M100 to be 
31.55 mag, i.e., a distance of 20.42 Mpc. Our main results as well as the
error contributions from various sources that were analyzed 
are summarized in Table~\ref{tab:err}.
The random error is  computed by finding the change in the intercept
$\mu$ when $\chi_1$ is increased by half the maximum deviation from the
best fit line after neglecting the three worst points.

Thus, our estimation of the distance to M100 is $20.42 \pm 1.7$ (random)
$\pm 2.4$ (systematic) Mpc. Taking into account the position of M100 relative
to the center of Virgo Cluster, the distance to the Virgo Center is
estimated to be $20.42 \pm 1.7$ (random) $\pm 2.6$ (systematic) Mpc.

The present work does not address the problem of recession velocity of the 
Virgo Center with respect to the Local Group. We would however, point out that
if one takes a central line-of-sight velocity dispersion, $\sigma_V$, of the 
order of 800 km~s$^{-1}$ and structural length, $a$ of 1.5 Mpc for the Virgo 
Cluster, the velocity of the Local Group towards Virgo produced by
the mass centered at Virgo Cluster would be of the order of
\beq
V_{\mathrm{peculiar}} \sim \frac{3 \sigma_V^2 a}{R^2}\, \tau 
\sim 75~\mathrm{km\,s}^{-1}
\enq
where $R$ is the distance to Virgo Center and $\tau$ is the age since the
formation of Virgo Cluster. Rowan-Robinson (1988) argued that
from the IRAS data there is no evidence for appreciable Virgo-centric
flow which is consistent with our simplistic calculation.
We take the recession velocity of Virgo to be $1170 \pm 80$ km s$^{-1}$
(cf.\ \cite{jt:93}) and estimate the Hubble Constant to be
\beq
H_0 = 57 \pm 5~\mathrm{(random)} \pm 8~\mathrm{(systematic)}~ 
\mathrm{km\,s}^{-1}\,\mathrm{Mpc}^{-1}
\enq

\begin{table}[!h]
\footnotesize
\caption{Results and Error Budget
\label{tab:err}}
\vskip 11pt
\begin{tabular}{lccccl}
\hline
\hline
\noalign{\vskip 8pt}
\multicolumn{1}{l}{Quantity} &
{Unit} &
{Mean} &
{Random} &
{Systematic} &
\multicolumn{1}{l}{Comments} \\
{} &
{} &
{Value} &
{Error} &
{Error} &
{}\\
\noalign{\vskip 4pt}
\hline
\noalign{\vskip 8pt}
\multicolumn{6}{l}{Calibration of the Period-Luminosity relation:} \\
Slope     & & $-3.45$ &      & 0.08 & No good test sample. \\
Intercept & mag & $-0.79$ &      & 0.20 & Distance to nearby Cepheids. \\
          & &       &      &      & Extinction correction. \\
\hline
\multicolumn{6}{l}{Period--$V$-magnitude relation for M100:} \\
Slope     & & $-3.49$ &      &      & Some Cepheids of $P > 55$ days\\
          & &       &      & \raisebox{-10pt}{0.10}
                                & could appear at 48--55 days. \\
Intercept & mag & \phs 30.80 & 0.10 &      & Extinction and incompleteness\\
          & &       &      &      & corrections not independent. \\
\hline
Extinction& mag & \phs 0.30 & 0.08 & 0.08 & Error in $\vi $ large \\
correction& &       &      &      & Reddening due to unresolved stars.   \\
          & &       &      &      & Recession of M100 --- K correction. \\
          & &       &      &      & Galactic \pca\ \\
          & &       &      &      & relations not well-determined. \\
\hline
Incompleteness& mag & \phs 0.28 & 0.12 &  & Model for the efficiency of \\
correction&   &     &      &      & detection not known. \\
          &   &     &      &      & Error in periods. \\
\hline
Zero point&  mag  &     &   & 0.08 & Observational problem. \\
calibration of&  &     &         &      &             \\
the detector& &  &      &      &             \\
\hline
\hline
Distance modulus  & mag &  31.55 & 0.18 & 0.26 &             \\
to M100           &     &        &      &      &             \\
Distance to M100  & Mpc &  20.42 &  1.7 &  2.4 &             \\
Distance to Virgo & Mpc &  20.42 &  1.7 &  2.6 & Position of M100 with \\
\tablenotemark{ } &     &        &      &      & respect to Virgo center. \\
\hline
Recession velocity&km/s & 1170   &      &  80  & Infall to Virgo Center of \\
of Virgo          &     &        &      &      & Local Group not same as  \\
                  &     &        &      &      & velocity component \\
                  &     &        &      &      & towards Virgo.\\
\hline
Hubble Constant   & km/s/Mpc & 57 &  5  &   8  &             \\
\noalign{\vskip 5pt}
\hline
\end{tabular}
\end{table}

\section{Summary and  Prospects}
\label{sec:summary}

Our strategy to investigate the calibration of Cepheids based on 
extragalactic distance scale is two-fold:
\bei
\item
Preparation of a reasonably well tested local complete sample of the parent 
population of Cepheid variables and quantification of some of
their characteristics for using them as benchmarks for
a determination of distance to far off galaxies.
\item
Carrying out tests on a set of homogeneous data of good quality for 
the external galaxy
and devise a method to extract the calibration characteristics without
getting unduly distorted by the noise.
\eni

For the Galactic Cepheid variables, we were guided by the \lig s,
number density as function of period and amplitude, and by the theory of
stellar pulsation. 
Our attempts to determine the position of Galactic Cepheids in the 
{\em surface temperature--surface gravity plane} was not very successful:
we are not very confident with our comparison of
the theoretical colors with the observed values
because the presently available
model atmospheric $\UB$ calibration appears to need better
input physics. The models based on stellar pulsation are limited by
the size of the helium core, and the boundary conditions in the
outer envelope where convection is supersonic, apart from the more
fundamental problem of coupling between convection and pulsation.
Still, as a working model, the \gteff\ strip we have determined
and the \pca\ relations we obtained should be
useful for the calibration of the Cepheid distance scale.

We used HST data on Cepheids for galaxies at distance modulus of the order
to 28 to 29 magnitudes to determine the slope of the period--$V$-magnitude
relation for the population that would be targeted for the measurement of
distances to farther galaxies.
The slope of the relation appears to be consistent between various galaxies.
Equally important is the intrinsic scatter in the relation
if we wish to provide a trustworthy error analysis of our
distance estimations, but we do not have good enough data yet to determine the
extinction corrected scatter. We have discussed the implications of
this drawback.

We have used $L_1$ minimization for the determination of distance modulus
of M100. We have attempted a correction for flux-limited incompleteness
by using a diagram of the relative number density of Cepheids in M100
as function of the period. We also carried out numerical simulations
using a toy model for the distribution function of the population
and the efficiency of the
detector. We provide a prescription for correction to  offset
the flux-limited incompleteness in a sample when a volume-limited
test sample of the same population is available.

It is indicated from our analysis that a reliable estimate of the
distance to galaxies situated within 40 Mpc is well within the capability
of the HST provided the observing strategy addresses some of the problems
specific to Cepheids which we have attempted to highlight in the present work.
But it should also be realized that a systematic error of approximately 
0.25 magnitude should be eliminated by observing a selected sample of
local Cepheid variables.

\acknowledgements

We are grateful to W. Freedman for sending the LMC data of Cepheid
variables and to L. Ferrarese for sending a draft of their work
on the HST key project on Cepheids in M100. We are thankful to S.M.Chitre for
many helpful comments on the manuscript. We acknowledge support from 
the Indo-French Center for the Promotion of Advanced Research 
(Project 1410-2).

\newpage
\bec
APPENDIX
\enc
\appendix

\section{Incompleteness Correction: Mathematical Formulation 
 and discussion of certain Systematics}
\label{app:incomp}

The astrophysical data is very often prone to incompleteness due to
limitation in the flux of radiation received from the source, thereby
systematically favoring detection of the brighter of the objects
having otherwise nearly identical properties. This effect, known as 
Malmquist bias, has been discussed extensively in the literature 
(e.g.\ \cite{sandage:87}). Even though theoretically this effect is pretty
well understood, its quantification is nontrivial because the local 
volume-limited sample is usually subject to large random and systematic errors
due to small number statistics or specific environment effects, though
it is not limited by the faintness of the objects. 
Thus, for example, the local environments of Cepheid variables in our
neighborhood may be different from that of a distant galaxy we are
observing and hence the local sample may not represent the parent
population to be analyzed or there may be too few stars within
some range of period in our complete sample causing the random noise
to overwhelm the properties we intend to characterize. 
Consequently, researchers are rarely in
agreement on whether a given set of data is biased due to flux limitation
or whether a correction needs to be applied to offset the Malmquist bias. 
In this section, we  try to
use a simple mathematical model to estimate the effect and later provide
an alternative easy-to-implement way using a diagram which can
be drawn with the available data, provided the observations of the
local sample and of the distant objects are carried out with a few precautions.
By comparing these results, we give a prescription for the quantification
of Malmquist bias in a sample, with the Cepheid population as a
specific example.

For the sake of tractability and ease of interpretation, we analyze
a simple model for the unnormalized probability density as
a function  of the $V$-magnitude
for a constant period of the Cepheid variables, of the form
\beq
f(V,P_{\mathrm{tr}}) = N(P_{\mathrm{tr}}) \, \exp \left[ 
    - \frac{\{ V- \mu + \alpha P_{\mathrm{tr}} \} ^2}{2 \sigma^2} \right],
\enq
where $V$ is the extinction corrected mean $V$-magnitude of the star,
$P_{\mathrm{tr}}$ is $\lp $ of pulsation if there were no error in
the estimation of the period,
$\mu $ is the zero point of the Cepheid period--$V$-magnitude relation for the
galaxy, $\alpha $ the negative of the slope, and
$\sigma $ the intrinsic scatter in the relation.
The expression for the normalization term $N(P_{\mathrm{tr}})$ 
in the probability density could have explicit dependency on period as well as
V-magnitude, but ours seemed to be a reasonable approximation.
The distribution of Cepheid variables within the instability strip is 
far from Gaussian, but that deviation will only increase the incompleteness and
consequently, in our first attempt to study this systematic effect we will not
be overestimating the correction if we use the Gaussian form.
The errors in the observed period is a major handicap, specifically to
estimate the magnitude at which the incompleteness becomes important.
But we take the simplistic view that the  probability density can be expressed
as a function of the observed  period by suitably weighted integration of the
above expression over period, and that the only change due to the integration is
an increase in the scatter $\sigma$ as argued in Section~\ref{sec:incomp}.
Further, depending upon whether extinction correction is carried out
independent of the incompleteness correction or not, the value of an 
effective $\sigma_{\mathrm{eff}}$ will be defined to incorporate the scatter 
in the observed period vs $V$-magnitude diagram after the extinction 
correction is over.
We make the working hypothesis that the efficiency of the
detector to find a Cepheid variable depends only on the apparent magnitude of
the star, $V_a$, and consequently we can analyze the incompleteness correction
at a fixed period of pulsation. Accordingly, we assume that the efficiency
of detection, $D(V_a)$ is given by
\beq
D(V_a) = 
   \cases{
 1 & if $ V_a \leq V_{0}$ \cr
 \exp [- \gamma (V_{0} - V_a)/\alpha ] & if $ V_a > V_{0}$ 
}
\enq
where $V_{0}$ is the cut off magnitude below which there is no incompleteness
of the sample due to flux limitation.

On account of flux limitation, the distribution function at the
observed period $P_{0} \equiv \log (P/{\mathrm{day}}) $ becomes
\beq
f_{0}(V_a, P_{0}) = f(V, P_{0}) \, D(V_a) 
\enq
We assume that subject to detectability, all extinction values $> 0$ are 
allowed. Hence, integrating over the extinction, $\epsilon =
V_a - V$ from 0 to $\infty $ at equal weight, 
we get the probability density within a 
normalization constant, as
\beq
\Theta(V,P_{0}) = 
\cases{
        [ 1 + \gamma (V_{0} - V)/\alpha]\, f(V,P_{0}) & if $V \leq V_{0}$ \cr
        \exp [\gamma (V_{0} - V)/\alpha]\, f(V,P_{0}) & if $V > V_{0}$
}
\enq
The probability density can thus be written as
\beq
F_{obs}(V,P_{0}) = \frac{\Theta(V,P_{0})}
{\displaystyle{\int \Theta(V,P_{0}) \, dV }} 
\enq
If there is no incompleteness, the expectation value of
$(V - \mu + \alpha P_{0})$ will be zero. The flux limitation decreases the 
value and this decrement is a measure of the incompleteness correction.

When $(\mu - \alpha P) \ll V_{0}$, the incompleteness will be negligible
and if $(\mu - \alpha P -V_{0}) > 2 \sigma_{\mathrm{eff}}$, the incompleteness
will tend to a constant value $(= \gamma  \sigma_{\mathrm{eff}}^2/\alpha$).
Results from this model calculations for various values of
$V_{0}$ and $\sigma_{\mathrm{eff}}$ are given in Table~\ref{tab:incomp}.

The various simplistic approximations used render the error analysis rather
difficult. But most of the difficulties could be resolved if we use
the observed data of the complete sample of local Cepheids and the
stars in the external galaxy.  The incompleteness could be reliably estimated
by resorting to the equivalent of Figure~\ref{fig:m100_ratio} with
the $V$-magnitude as the abscissa, if the ratio of the number
density as a function of $V$-magnitude (i.e. the variable determining the 
incompleteness) had been available. In the absence of
such a plot, the period of pulsation is converted into an
equivalent $V$-magnitude and used for the incompleteness correction
through Figure~\ref{fig:m100_ratio}. If indeed we can use the
$V$-magnitude as abscissa, the measure of incompleteness at a specified
period of pulsation is simply the {\em scatter in the $V$-magnitude for a 
fixed period, $ \sigma_{\mathrm{eff}}$, times the
decrease in the logarithm of the ratio of the number density in the
observed sample to the number density in the complete sample
when the abscissa is increased by an amount $\sigma_{\mathrm{eff}}$}.
Since these quantities can be plotted without any detailed modeling
like we have shown in Figure~\ref{fig:m100_ratio}, 
we can carry out the extinction correction as well as possible error 
analysis in our estimate according to the approximate prescription given in 
Section~\ref{sec:incomp} and the numerical values are shown in 
Table~\ref{tab:incomp}.
But it should be stressed that, though the dip in the observed number density
of Cepheids in M100 at period of 35 to 45 days due to detection strategy
does not directly introduce incompleteness corrections, 
it makes it difficult to determine
the cutoff magnitude $V_{0}$ beyond which there is flux-limited incompleteness.
Consequently, the peak of the flat region in Figure~\ref{fig:m100_ratio} 
as well as the period
at which the peak is attained are uncertain. This is the region where
our approximation based on Figure~\ref{fig:m100_ratio} and the numerical 
simulations differ
considerably as we see in Table~\ref{tab:incomp}, 
but since both the methods are subject to systematic errors
introduced because of the distortion in the figure, we do not feel either
of the method is superior.

{}


\begin{thebibliography}{}

\bibitem[Barrodale \& Zala 1986]{bz:86} Barrodale, I. \& Zala, C. 
        1986, in Numerical Algorithms, ed. J. L. Mohamed \& J. E. Walsh
        (Clarendon Press, Oxford), 220
\bibitem[Bhat, Gandhi \& Narasimha 1998]{bgn:98} Bhat, M., Gandhi, B., \&
        Narasimha, D.
        1998, work in progress
\bibitem[Cardelli, Clayton \& Mathis 1989]{ccm:89} Cardelli, J. A.,
        Clayton, G. C., \& Mathis, J. S.
        1989, \apj, 345, 245
\bibitem[Feast \& Catchpole 1997]{fc:97} Feast, M. W., \& Catchpole, R. M.
        1997, \mnras, 286, L1
\bibitem[Ferrarese \ea 1996]{ferr:96} Ferrarese, L., \ea
        1996, \apj, 464, 568
\bibitem[Freedman \ea 1994]{freedman:94} Freedman, W. L., \ea
        1994, \nat, 371, 757
\bibitem[Freedman 1995]{freedman:95} Freedman, W. L.
        1995, private communication
\bibitem[Freedman \ea 1998]{freedman:98} Freedman, W. L., Mould, J. R.,
        Kennicutt, R. C., Madore, B. F.
        1998, in IAU Symposium No. 183, Cosmological 
        Parameters and the Evolution of the Universe, ed. K. Sato, in
        preparation (astro-ph/9801080)
\bibitem[Graham \ea 1997]{graham:97} Graham, J. A., \ea
        1997, \apj, 477, 535
\bibitem[Horne \& Baliunas 1986]{hb:86} Horne, J. H., \& Baliunas, S. L.
        1986, \apj, 302, 757
\bibitem[Jerjen \& Tammann 1993]{jt:93} Jerjen, H., \& Tammann, G. A.
        1993, \aap, 276, 1
\bibitem[Morgan 1994]{morgan:94} Morgan, S.
        1994, ASP Conference Series, Vol. 57, 145
\bibitem[Press \& Rybicki 1989]{pr:89} Press, W. H., \&  Rybicki, G. B.
        1989, \apj, 338, 277
\bibitem[Rowan-Robinson 1988]{rowan:88} Rowan-Robinson, M.
        1988, \ssr, 48, 1
\bibitem[Saha \& Hoessel 1990]{sh:90} Saha, A., \& Hoessel, J. G.
        1990, \aj, 99, 97
\bibitem[Saha \ea  1994]{saha:94} Saha, A., Labhardt, L., Schwengeler, H.,
        Macchetto, F. D., Panagia, N., Sandage, A., Tammann, G. A.
        1994, \apj, 425, 14 
\bibitem[Saha \ea 1996]{saha:96} Saha, A., Sandage, A., Labhardt, L.,
        Tammann, G. A., Macchetto, F. D., Panagia, N.
        1996, \apj, 466, 55
\bibitem[Sandage 1987]{sandage:87} Sandage, A.
        1987, in IAU Symposium No. 124, Observational Cosmology, ed.
        A. Hewitt, G. Burbidge, \& L. Z. Fang, 1
\bibitem[Sandage \ea  1994]{sandage:94} Sandage, A., Saha, A., Tammann, G. A.,
        Labhardt, L., Schwengeler, H., Panagia, N., Macchetto, F. D.
        1994, \apj, 423, L13 
\bibitem[Silbermann \ea 1996]{silbermann:96} Silbermann, N. A., \ea
        1996, \apj, 470, 1
\end{thebibliography}
\end{document}